\documentclass[aip,jcp,reprint]{revtex4-1}
\usepackage{epsfig}
\usepackage{color}
\usepackage{amsmath}
\usepackage{amsfonts}
\usepackage{natbib}
\usepackage{notes2bib}
\usepackage{multirow}

\newcommand{\erf}{\ensuremath{\mathop{\rm erf}}}
\newcommand{\erfc}{\ensuremath{\mathop{\rm erfc}}}
\newcommand{\rb}{\mathbf{r}}

\newcommand{\len}[1]{\left|#1\right|}

\newcommand{\curly}[1]{\left\{#1\right\}}
\newcommand{\para}[1]{\left(#1\right)}

\renewcommand{\vec}[1]{\mathbf{#1}}

\newcommand{\kT}{\ensuremath{k_{\rm B}T}}

\newcommand{\pvec}[1]{\vec{#1}\mkern2mu\vphantom{#1}}

\begin{document}


\title{Self-Consistent Determination of Long-Range Electrostatics in Neural Network Potentials}



\author{Ang Gao}
\email[]{anggao@bupt.edu.cn}
\affiliation{Department of Physics, Beijing University of Posts and Telecommunications, Beijing, China 100876}
\author{Richard C. Remsing}
\email[]{rick.remsing@rutgers.edu}
\affiliation{Department of Chemistry and Chemical Biology, Rutgers University, Piscataway, NJ, USA 08854}




\begin{abstract}
Machine learning has the potential to revolutionize the field of molecular simulation
through the development of efficient and accurate models of interatomic interactions. 
In particular, neural network models can describe interactions at the level of accuracy of quantum mechanics-based calculations, but with a fraction of the cost,
enabling the simulation of large systems over long timescales with ab initio accuracy. 
However, implicit in the construction of neural network potentials is an assumption of locality, wherein atomic arrangements on the scale of about a nanometer are used to learn interatomic interactions. 
Because of this assumption, the resulting neural network models cannot describe long-range interactions that play critical roles in dielectric screening and chemical reactivity. 
To address this issue, we introduce the self-consistent field neural network (SCFNN) model --- a general approach for learning the long-range response of molecular systems in neural network potentials.   
The SCFNN model relies on a physically meaningful separation of the interatomic interactions into short- and long-range components, with a separate network to handle each component. 
We demonstrate the success of the SCFNN approach in modeling the dielectric properties of bulk liquid water, and show that the SCFNN model accurately predicts long-range polarization correlations and the response of water to applied electrostatic fields. 
Importantly, because of the separation of interactions inherent in our approach, the SCFNN model can be combined with many existing approaches for building neural network potentials. 
Therefore, we expect the SCFNN model to facilitate the proper description of long-range interactions in a wide-variety of machine learning-based force fields.
\end{abstract}


\maketitle

\raggedbottom

Computer simulations have transformed our understanding of molecular systems
by providing atomic-level insights phenomena of wide importance. 
The earliest models used efficient empirical descriptions of interatomic interactions, and similar force field-based simulations form the foundation of molecular simulations today~\cite{allen2017computer}. 
However, it is difficult to describe processes like chemical reactions that involve bond breakage and formation, as well as electronic polarization effects within empirical force fields. 
The development of quantum mechanics-based ab initio simulations enabled the description of these complex processes, leading to profound insights across scientific disciplines~\cite{tuckerman1996ab,car1985unified,chen2018hydroxide,geissler2001autoionization,lee2008role,walker2008implementation,senn2006qm,dal2007investigating}. 
The vast majority of these first principles approaches rely on density functional theory (DFT), and the development of increasingly accurate density functionals has greatly improved the reliability of ab initio predictions~\cite{sun2015strongly,Sun:NatChem:2016,Remsing:PNAS:2017,furness2020accurate,zhang2021modeling,adamo1999toward}. 
But, performing electronic structure calculations are expensive, and first principles simulations are limited to small system sizes and short time scales.

The prohibitive expense of ab initio simulations can be overcome through machine learning. 
Armed with a set of ab initio data, machine learning can be used to train neural network (NN) potentials that describe interatomic interactions at the same level of accuracy as the ab initio methods, but with a fraction of the cost. 
Consequently, NN potentials enable ab initio quality simulations to reach the large system sizes and long time scales needed to model complex phenomena, such as phase diagrams~\cite{gartner2020signatures,zhang2021phase,deringer2018realistic,niu2020ab,Deringer:2021aa} and nucleation~\cite{khaliullin2011nucleation,bonati2018silicon}. 

Despite the significant advances made in this area, there are still practical and conceptual difficulties with NN potential development, especially with regard to long-range electrostatics. 
To make NN potential construction computationally feasible, most approaches learn only local arrangements of atoms around a central particle, where the meaning of ``local'' is defined by a distance cutoff usually less than 1~nm. 
Because of this locality, the resulting NN potentials are inherently short-ranged. 
The lack of long-range interactions in NN potentials can lead to both quantitative and qualitative errors, especially when describing polar and charged species~\cite{Yue2021, Grisafi2019, niblett2021learning}. 

The need for incorporating long-range electrostatics into NN potentials has led to the development of several new approaches~\cite{Xie2020, Yao2018, Grisafi2019, Yue2021, Ko2021}.
Many of these approaches exclude all or some of the electrostatic interactions from training and then assign effective partial charges to each atomic nucleus that are used to calculate long-range electrostatic interactions using traditional methods~\cite{Xie2020, Yao2018, Yue2021, Ko2021, niblett2021learning}.
The values of these effective charges can be determined using machine learning methods. 
For example, 4G-HDNNP~\cite{Ko2021} employs deep neural networks to predict the electronegativities of each nucleus, which are subsequently used within a charge equilibration process to determine the effective charges.
These approaches can predict binding energies and charge transfer between molecules, but they also introduce quantities that are not direct physical observables, such as the effective charges and electronegativities. 
Another approach introduced feature functions to explicitly incorporate nonlocal geometric information into the construction of NN potentials~\cite{Grisafi2019}. 
However, these feature functions depend on the system size. 
The resulting NN models cannot be used to accurately model systems that are larger than the original training set. 
This size restriction severely limits applicability by making this approach unable to model extended system sizes. 

The difficulties that current approaches to NN potentials have when treating long-range interactions can be resolved by a purely ab initio strategy that uses no effective quantities. 
Such a strategy can be informed by our understanding of the roles of short- and long-range interactions in condensed phases~\cite{WidomScience,WCA,WCA-Science,LMFDeriv}. 
In uniform liquids, appropriately chosen uniformly slowly-varying components of the long-range forces --- van der Waals attractions and long-range Coulomb interactions --- cancel to a good approximation in every relevant configuration. 
As a result, the local structure is determined almost entirely by short-range interactions. 
In water, these short-range interactions correspond to hydrogen-bonding and packing~\cite{rodgers2009accurate,Remsing:JSP:2011,Rodgers:PNAS:2008,MolPhysLMF}. 
Therefore, short-range models, including current NN potentials, can describe the structure of uniform systems.
This idea, that short-range forces determine the structure of uniform systems, forms the foundation for the modern theory of bulk liquids~\cite{WidomScience,WCA,WCA-Science}, in which the averaged effects of long-range interactions can be treated as a small correction to the purely short-range system.

In contrast, the effects of long-range interactions are more subtle and play a role in collective effects that are important for dielectric screening. 
Moreover, long-range forces do not cancel at extended interfaces and instead play a key role in interfacial physics.
As a result, short-range systems cannot describe interfacial structure and thermodynamics, as they do in the bulk,
and standard NN models fail to describe even the simplest liquid-vapor interfaces~\cite{niblett2021learning}. 
The local molecular field (LMF) theory of Weeks and coworkers provides a framework for capturing the average effects of long-range interactions at interfaces through an effective external field~\cite{LMFDeriv,Remsing:2016ib,gao2018role,SSM,Remsing23874}.
LMF theory also provides physically intuitive insights into the roles of short- and long-range forces at interfaces that can be leveraged to model nonuniform systems.

Here, we exploit the physical picture provided by liquid-state theory to develop a general approach for learning long-range interactions in NN potentials from ab initio calculations. 
We separate the atomic interactions into appropriate short-range and long-range components and construct a separate network to handle each part. 
Importantly, the short-range model is isolated from the long-range interactions, such that each component is treated independently. 
This separation also isolates the long-range response of the system, enabling it to be learned. 
Short-range interactions can be learned using established approaches. 
The short- and long-range components of the potential are then connected through a rapidly-converging self-consistent loop. 
The resulting self-consistent field neural network (SCFNN) model is able to describe the effects of long-range interactions without the use of effective charges or similar artificial quantities. 
We illustrate this point through the development of a SCFNN model of liquid water. 
In addition to capturing the local structure of liquid water, as evidenced by the radial distribution function, the SCFNN model accurately describes long-range structural correlations connected to dielectric screening, as well as the response of liquid water to electrostatic fields.

\section{Results}

\subsection{Workflow of the Self-Consistent Field Neural Network Model}

\begin{figure*}[tb]
\centering
\includegraphics[scale=0.3]{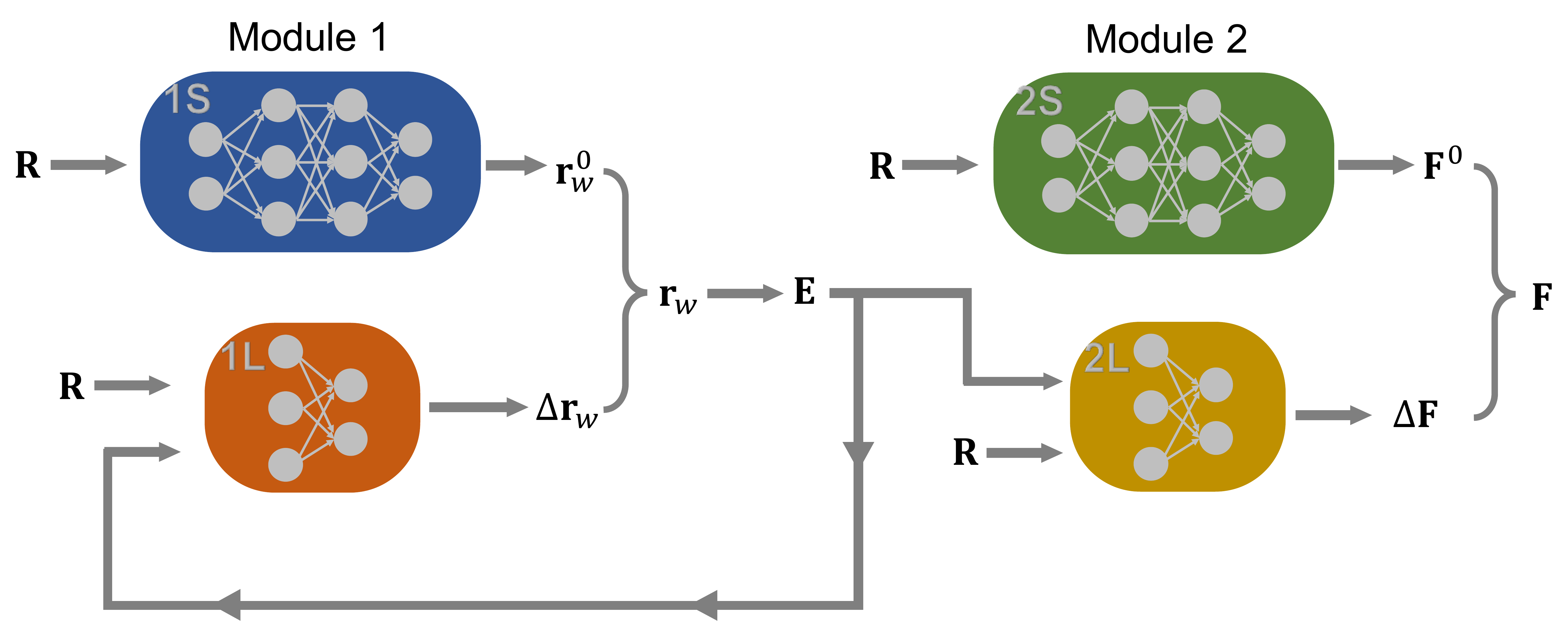}
\caption{Schematic of the self-consistent field neural network (SCFNN).
The SCFNN consists of two modules, each with two networks.
One networks learns the short-range interactions (S) and the other learns the effects of long-range interactions (L). 
Module 1 learns the positions of maximally localized Wannier function centers, $\vec{r}_w$, and Module 2 learns the forces, $\vec{F}$, on the atomic nuclei, the positions of which are indicated by $\vec{R}$.}
\label{fig:workflow}
\end{figure*}

The SCFNN model consists of two 
modules that each target a specific response of the system (Fig.~\ref{fig:workflow}). 
Module 1 predicts the electronic response via the position of the maximally localized Wannier function centers (MLWFCs). 
Module 2 predicts the forces on the nuclear sites. 
In turn, each module consists of two networks: one to describe the short-range interactions
and one to describe perturbations to the short-range system from long-range electric fields. 
Together, these two modules (four networks) enable the model to predict the total electrostatic properties of the system. 
In the short-range system, the $v(r)=1/r$ portion of the Coulomb potential is replaced by the short-range potential $v_0(r)=\erfc(r/\sigma)/r$.
Physically, $v_0(r)$ corresponds to screening the charge distributions in the system through the addition of neutralizing Gaussian charge distributions of opposite sign - the interactions are truncated by Gaussians.  
Therefore, we refer to this system as the Gaussian-truncated (GT) system~\cite{Rodgers:PNAS:2008,Remsing:JSP:2011,rodgers2009accurate,MolPhysLMF,LMFDeriv}.
By making a physically meaningful choice for $\sigma$, the GT system can describe the structure
of bulk liquids with high accuracy but with a fraction of the computational cost. 
Moreover, the GT system has served as a useful short-range component system when modeling the effects of long-range fields~\cite{Rodgers:PNAS:2008,Remsing:2016ib,SSM,baker2020local,Cox}. 
Here, we choose $\sigma$ to be 4.2~\AA \ (8 Bohr), which is large enough for the GT system to accurately describe hydrogen bonding and the local structure of liquid water~\cite{Rodgers:PNAS:2008,Remsing:JSP:2011,rodgers2009accurate,MolPhysLMF,LMFDeriv}.

%
The remaining part of the Coulomb interaction, $v_1(r)=v(r)-v_0(r) = \erf(r/\sigma)/r$, is long ranged, but varies slowly over the scale of $\sigma$.
Because $v_1(r)$ is uniformly slowly-varying, the effective field produced by $v_1(r)$ usually induces a linear response in the GT system. 
The linear nature of the response makes the effects of $v_1(r)$ able to captured by linear models. 
In the context of neural networks, we demonstrate below that a linear network is sufficient to learn the linear response induced by long-range interactions. 

\subsubsection{Module 1}
\begin{figure}[tb]
\centering
\includegraphics[scale=0.3]{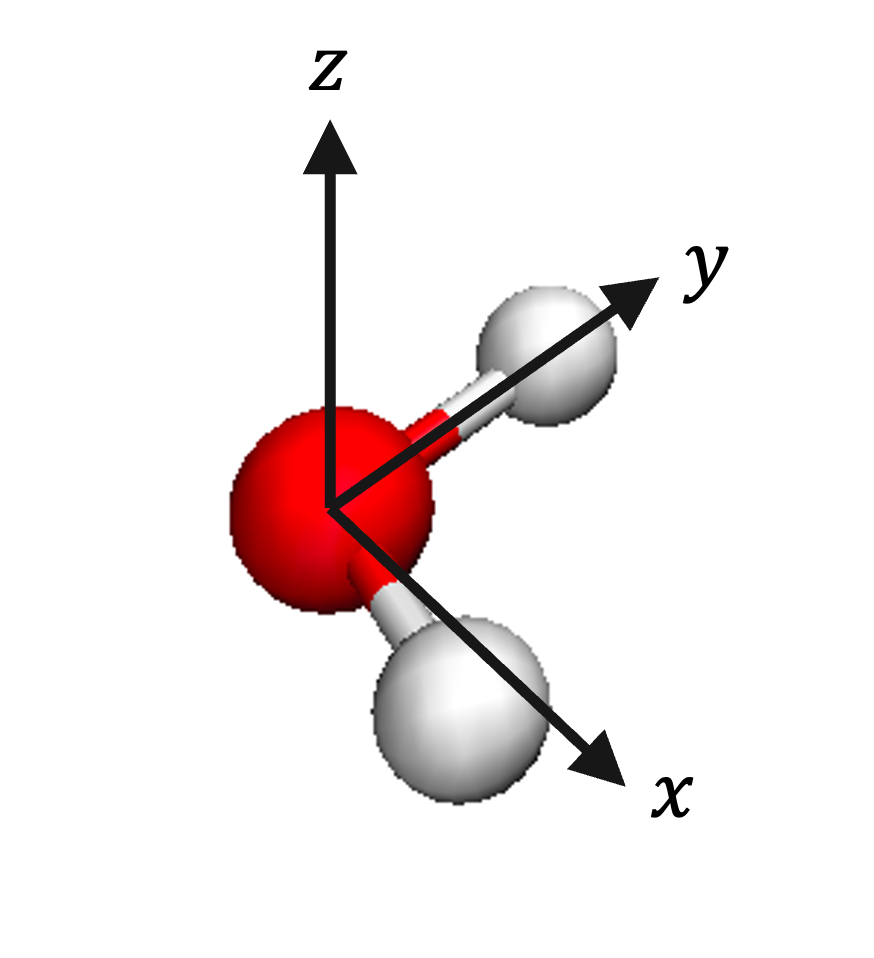}
\caption{Local frame around the central water. The $y$-axis is along the OH bond. The $z$-axis is perpendicular to the plane of the molecule. The $x$-axis is perpendicular to the $x$ and $z$ axis.}
\label{fig:localframe}
\end{figure}

The separation of interactions into short- and long-range components is crucial to the SCFNN model. 
In particular, the two networks of each module are used to handle this separation. 
Network 1S of Module 1 predicts the positions of the MLWFCs in the short-range GT system,
while Network 1L predicts the perturbations to the MLWFC positions induced by the effective long-range field. 
 Networks 1S and 1L leverage Kohn's theory on the nearsightedness of electronic matter (NEM)~\cite{Kohn1996, Prodans2005}. 
 The NEM states that~\cite{Prodans2005} ``local electronic properties, such as the density $n(r)$, depend significantly on the effective external potential only at nearby points.''
Here, the effective external potential refers to the Kohn-Sham effective potential, which includes the external potential
and the self-consistently determined long-range electric fields. 
Therefore, the NEM suggests that the electronic density, and consequently the positions of the MLWFCs, are `nearsighted' with respect to the effective potential, but not to the atomic coordinates, contrary to what has been assumed in previous work that also uses local geometric information of atoms as input to neural networks~\cite{Krishnamoorthy2021,Zhang2020}. 
An atom located at $\rb'$ will affect the effective potential at $\rb$, even if $\rb'$ is far from $\rb$, through long-range electrostatic interactions. 
Consequently, current approaches to generating NN models can only predict the position of MLWFCs for a purely short-range system without long-range electrostatics, such as the GT system~\cite{Krishnamoorthy2021,Zhang2020}.
We exploit this fact and use established NNs to predict the locations of the MLWFCs in the GT system~\cite{Krishnamoorthy2021}.
To do so, we create a local reference frame around each water molecule (Fig.~\ref{fig:localframe}) and use the coordinates of the surrounding atoms as inputs to the neural network. 
The local reference system preserves the rotational and translational symmetry of the system.
The network outputs the positions of the four MLWFCs around the central water, which are then transformed to the laboratory frame of reference.

Network 1L predicts the response of the MLWFC positions to the effective field $\vec{E}(\vec{r})$,
defined as the sum of the external field, $\vec{E}_{\rm ext}(\vec{r})$, and the long-range field from $v_1(r)$:
 \begin{equation}
 \vec{E} (\vec{r})= \vec{E}_{\rm ext} (\vec{r}) + \int d \pvec{r}' \rho(\pvec{r}') \nabla v_1(|\vec{r} - \pvec{r}'|) \,,
 \end{equation} 
 where $\rho(\pvec{r}')$ is the \emph{instantaneous} charge density of the system, including nuclear and electronic charges. 
 Network 1L also introduces a local reference frame for each water molecule.
 However, Network 1L takes as input both the local coordinates \emph{and} local effective electric fields.
 The NEM suggests that this local information is sufficient to determine the perturbation in the MLWFC positions. 
 Network 1L outputs this change in the positions of the water molecule's four MLWFCs,
 and this perturbation is added to the MLWFC position determined in the GT system to obtain
 the MLWFCs in the full system. 
 We note that $\vec{E}(\vec{r})$ is a slowly-varying long-range field,
 such that the MLWFCs respond linearly to this field. 
 Therefore, Network 1L is constructed to be linear in $\vec{E}(\vec{r})$.
 Table~1 demonstrates that the linear response embodied by Network 1L predicts
 the perturbation of the MLWFCs with reasonable accuracy. 
 
%
We now need to determine the effective field $\vec{E}(\vec{r})$.
This effective field depends on the electron density distribution, but evaluating and including the full three-dimensional electron density for every configuration in a training set requires a prohibitively large amount of storage space. 
Instead, we approximate the electron density by the charge density of the MLWFCs, assuming each MLWFC is a point charge of magnitude $-2e_0$.
This approximation is often used when computing molecular multipoles, as needed to predict vibrational spectra, for example~\cite{Zhang2020,zhang2021modeling}. 
Here, it is important to note that the MLWFs of water are highly localized, so that the center gives a reasonable representation of the location of the MLWF. 
Moreover, the electron density is essentially smeared over the scale of $\sigma$ through a convolution with $v_1(r)$, which makes the resulting fields relatively insensitive to small-wavelength variations in the charge density. 
As a result, the electron density can be accurately approximated by the MLWFC charge density within our approach.

The effective field is a functional of the set of MLWFC positions, $\vec{E}[\curly{\vec{r}_w}]$, and the positions of the MLWFCs themselves depend on the field, $\vec{r}_w[\vec{E}]$. 
Therefore, we determine $\vec{E}$ and $\curly{\vec{r}_w}$ through self-consistent iteration.
Our initial guess for $\vec{E}$ is obtained from the positions of the MLWFCs in the GT system. 
We then iterate this self-consistent loop until 
the MLWFC positions no longer change,
within a tolerance of $2.6\times10^{-4}$~\AA.
In practice, we find that self-consistency is achieved quickly.

\subsubsection{Module 2}
After Module 1 predicts the positions of the MLWFCs, Module 2 predicts the forces on the atomic sites. 
As with the first module, Module 2 consists of two networks: one that predicts the forces of the GT system and another that predicts the forces produced by $\vec{E}(\vec{r})$.
To predict the forces in the GT system, we adopt the network used by Behler and coworkers~\cite{Morawietz2016}.
This network, Network 2S, takes local geometric information of the atoms as inputs and, consequently, cannot capture long-range interactions. 
To describe long-range interactions, we introduce a second network (Network 2L in Fig.~\ref{fig:workflow}). 
This additional network predicts the forces on atomic sites due to the effective field $\vec{E}(\vec{r})$, which properly accounts for long-range interactions in the system. 
In practice, we again introduce a local reference frame for each water molecule and use local atomic coordinates and local electric fields as inputs. 
In this case, we also find that a network that is linear in $\vec{E}(\vec{r})$ accurately predicts the resulting long-range forces, consistent with the linear response of the system to a slowly-varying field. 

In practice, separating the data obtained from standard DFT calculations into the GT system and the long-range effective field is not straightforward. 
To solve this problem, we apply homogeneous electric fields of varying strength while keeping the atomic coordinates fixed. 
The fields only perturb the positions of the MLWFCs and the forces on the atoms --- these perturbations are not related to the GT system. 
The changes induced by these electric fields are directly obtained from DFT calculations and are used to train Networks 1L and 2L, which learn the response to long-range effective fields. 
The remaining part of the DFT data is used to train Networks 1S and  2S, which learn the response of the short-ranged GT system. 
See the Methods section for a more detailed discussion of the networks and the training procedure.

We emphasize that our approach to partitioning the system into a short-range GT piece and a long-range perturbation piece is different from other machine learning approaches for handling long-range electrostatics. 
The standard approaches usually partition the total energy into two parts, a short-ranged energy and an Ewald energy that is used to evaluate the long-range interactions. 
However, this partitioning results in a coupling between the short- and long-range interactions.
For example, the short-range part of the energy in the 4G-HDNNP model depends on the effective charges that are assigned to the atoms, but these effective charges depend on long-range electrostatic interactions through the global charge equilibration process used to determine their values~\cite{Ko2021}.
In contrast, the approach we propose here isolates the short-range and long-range physics. 
The GT system does not depend on long-range electrostatics even implicitly; it is completely uncoupled from the long-range interactions. 
The effects of long-range electrostatic interactions are completely isolated within the second network of each module, Network 1L and Network 2L in Fig.~\ref{fig:workflow}. 
This separation of short- and long-ranged effects is similar in spirit to the principles underlying LMF theory~\cite{LMFDeriv,Remsing:2016ib,SSM} and related theories of uniform liquids~\cite{WCA,WCA-Science,MolPhysLMF,rodgers2009accurate,Remsing:JSP:2011}.

\begin{table}[tb]
\centering
\begin{tabular}{|c|c|c|c|c|c|c|}
\hline
 \multirow{2}{*}{ } & \multicolumn{3}{c|}{$0.1$~V/\AA} &  \multicolumn{3}{c|}{$0.2$~V/\AA} \\ \cline{2-7}
                          &  MLWFC & $F_{\rm O}$ & $F_{\rm H}$ &  MLWFC & $F_{\rm O}$ & $F_{\rm H}$ \\ \hline
                   MAE ($\times 100$) &$0.028 $  &$1.4 $ &$0.98 $&$0.056 $    &$2.8$  &$2.0$ \\ \hline
                 \end{tabular}
\label{table:accuracy_24}
\caption{When homogeneous external fields are applied, the location of the maximally localized Wanner function center (MLWFC) and the forces on the oxygen and hydrogen nuclei, $F_{\rm O}$ and $F_{\rm H}$, respectively, are changed.
This table shows the Mean Absolute Error (MAE) of Network 1L and 2L in predicting the changes in the MLWFC positions (\AA) and the  forces (eV/\AA) along the $z$-direction when fields of strength 0.1~V/\AA \ and 0.2~V/\AA \ are applied along $z$-direction. 
The predictions are made for the test sets and the error is computed with respect to the DFT results.}
\end{table}

\begin{figure}[tb]
\centering
\includegraphics[scale=0.8]{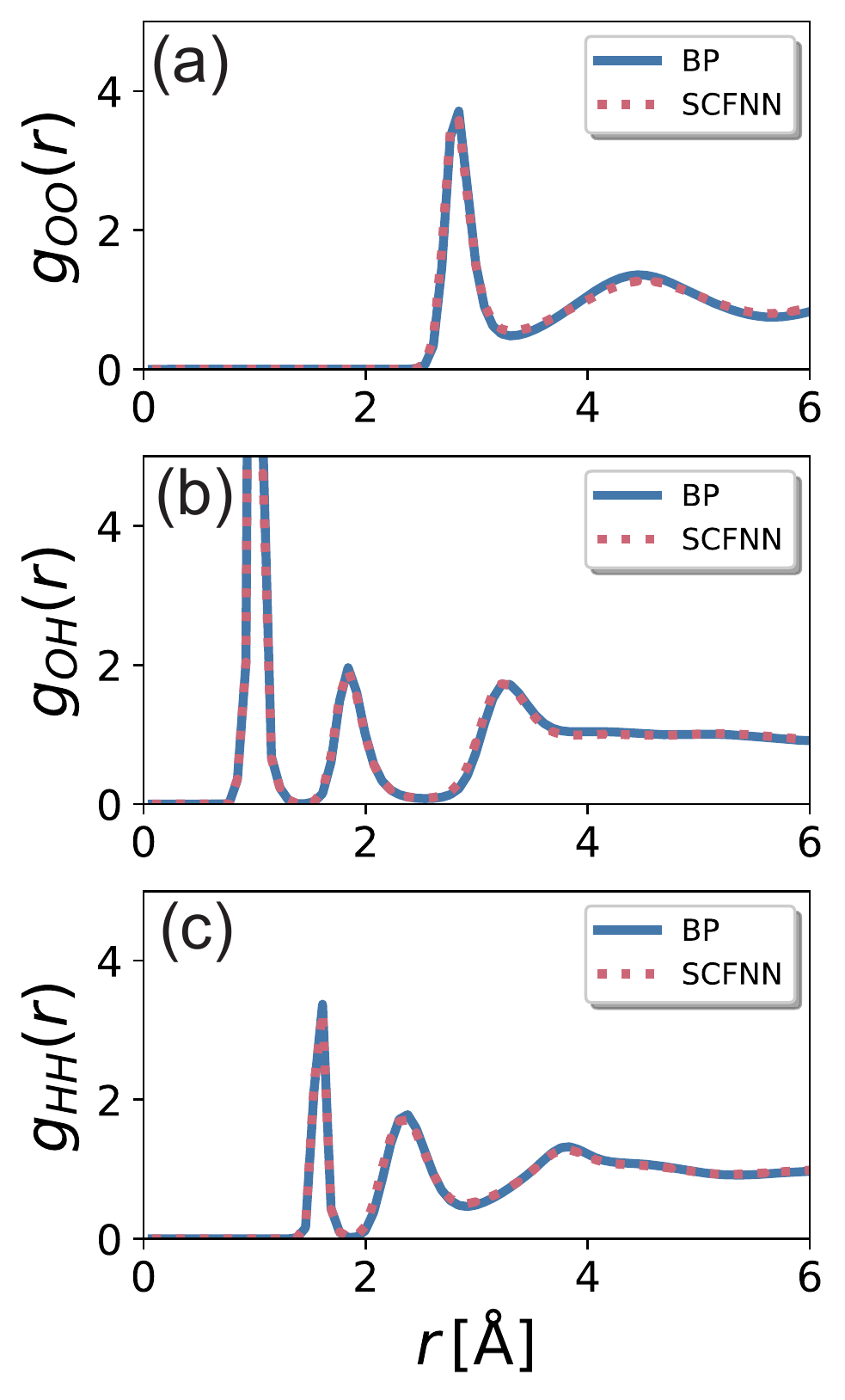}
\caption{Comparison of the radial distribution functions for (a) O-O, (b) O-H and (c) H-H correlations in liquid water,
as predicted by molecular dynamics simulations of the self-consistent field neural network (SCFNN) and Behler-Parrinello (BP) models.}
\label{fig:rdf}
\end{figure}

\subsection{Water's Local Structure is Insensitive to Long-Range Interactions}

We demonstrate the success of the SCFNN approach by modeling liquid water. 
Water is the most important liquid on Earth. 
Yet, the importance of both short- and long-range interactions makes it difficult to model. 
Short-range interactions are responsible for water's hydrogen bond network that is essential to its structure and unusual but important thermodynamic properties~\cite{Ball:2008aa,Remsing:JSP:2011}. 
Long-range interactions play key roles in water's dielectric response, interfacial structure, and can even influence water-mediated interactions~\cite{prelesnik2021ion}. 
Because of this broad importance, liquid water has served as a prototypical test system for many machine learning-based models~\cite{Morawietz2016,cheng2019ab,Zhang2020,zhang2021phase,Grisafi2019}
Here, we test our SCFNN model on a system of bulk liquid water by performing molecular dynamics simulations of 1000 molecules in the canonical ensemble under periodic boundary conditions.

One conventional test on the validity of a NN potential is to compare the radial distribution function, $g(r)$, between atomic sites for the different models. 
The $g(r)$ predicted by the SCFNN model is the same as that predicted by the Behler-Parrinello model~\cite{Morawietz2016} for all three site-site correlations in water (Fig.~\ref{fig:rdf}). 
This level of agreement may be expected, based on previous work examining the structure of bulk water~\cite{Rodgers:PNAS:2008,MolPhysLMF,Remsing:JSP:2011,gao2018role,SSM}. 
The radial distribution functions of water are determined mainly by short-range, nearest-neighbor interactions, which arise from packing and hydrogen bonding; long-range interactions have little effect on the main features of $g(r)$. 
Consequently, purely short-range models, like the GT system, can quantitatively reproduce the $g(r)$ of water~\cite{Rodgers:PNAS:2008,MolPhysLMF,Remsing:JSP:2011,gao2018role,SSM}. 
Similarly, the short-range Behler-Parrinello model accurately describes the radial distribution functions, as does the SCFNN model, which includes long-range interactions.

\subsection{Long-Range Electrostatics and Dielectric Response}

Though the short-range structure exemplified by the radial distribution function is insensitive to long-range interactions, long-range correlations are not. 
For example, the longitudinal component of the dipole density or polarization correlation function evaluated in reciprocal space, $\chi^0_{zz}(\vec{k}) $, was recently shown to be sensitive to long-range interactions~\cite{Cox}. 
This correlation function is defined according to
\begin{equation}
\chi^0_{zz}(\vec{k})  = \frac{1}{V} \sum_{l,j}\frac{(\vec{k} \cdot \vec{p}_l) \,(\vec{k}  \cdot \vec{p}_j) }{k^2}\, e^{-i\vec{k} \cdot \left(\vec{r}_l - \vec{r}_j \right) } \,,  \,\, \mathrm{with} \,\,\vec{k} = k\hat{\vec{z}}\,.
\end{equation}
Here $\vec{p}_j$ is the dipole moment of water molecule $j$ and $\vec{r}_j$ is the position of the oxygen atom of water molecule $j$.

\begin{figure}[tb]
\centering
\includegraphics[scale=0.6]{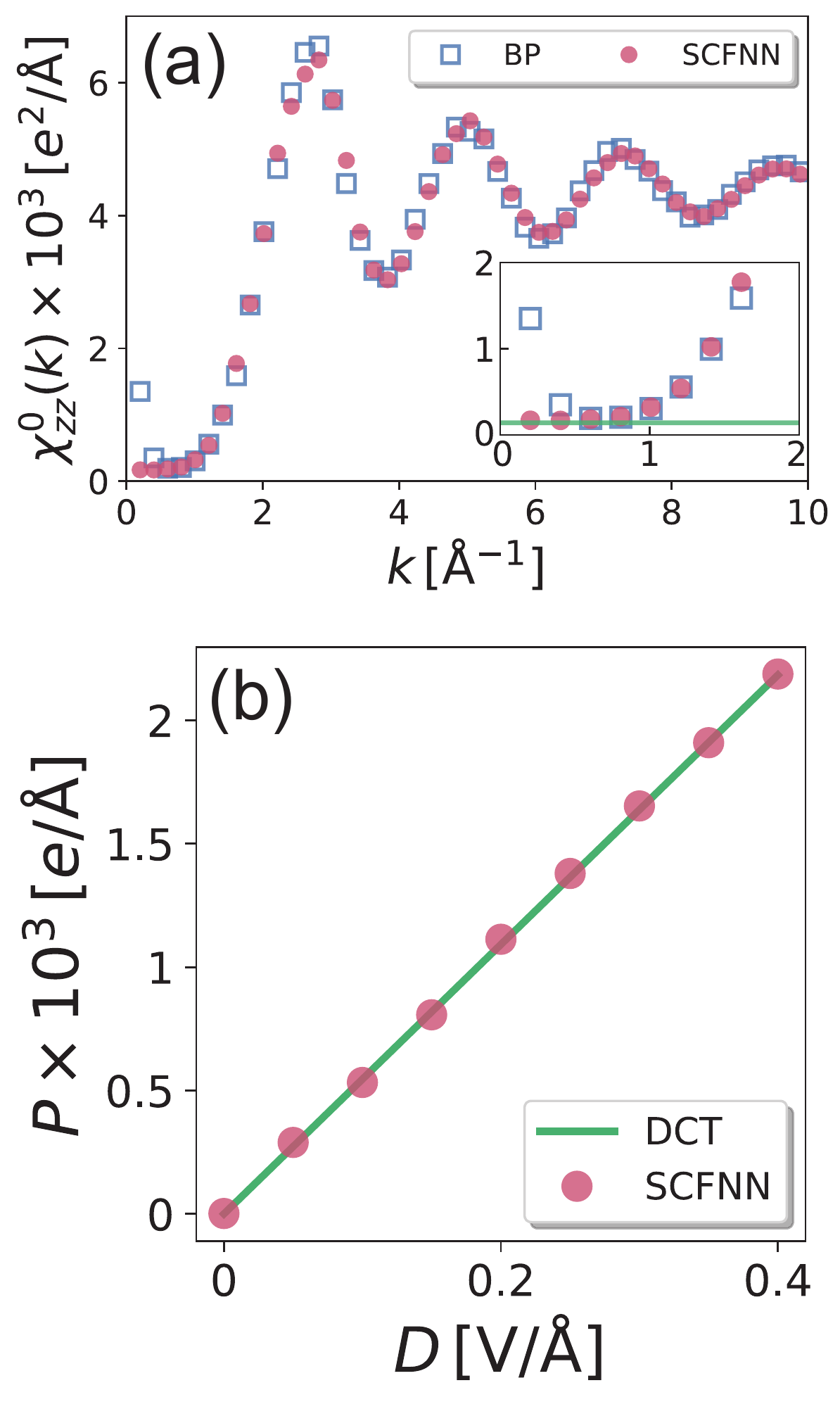}
\caption{(a) The longitudinal polarization correlation function in reciprocal space, $\chi^0_{zz}(\vec{k}) $,  shows differences between the self-consistent field neural network (SCFNN) and Behler-Parrinello (BP) models at low $k$. In particular, the SCFNN model plateaus as $k\rightarrow0$ in a manner consistent with the theoretical prediction (green line), while the BP (short-range) model does not. 
(b) The polarization, $P$, induced by a homogeneous displacement field of magnitude $D$ along the $z$-axis is accurately predicted by the SCFNN model, evidenced by the agreement with dielectric continuum theory (DCT) predictions.}
\label{fig:dipole}
\end{figure}

The longitudinal polarization correlation function predicted by our SCFNN model and the Behler-Parrinello agree everywhere except at small $k$, indicating that long-range correlations are different in the two models (Fig.~\ref{fig:dipole}a). 
The long-wavelength behavior of the polarization correlation function is related to the dielectric constant via
\begin{equation}
\lim_{k\rightarrow 0} \chi^0_{zz}(\vec{k})  = \epsilon_0 \kT \frac{\epsilon - 1}{\epsilon}\,,
\end{equation}
where $\epsilon=78.4$ is the dielectric constant of water.
The $\chi^0_{zz}(\vec{k}) $ predicted by our SCFNN model is consistent with the expected behavior at small $k$.
In contrast, short-range models, like the GT system~\cite{Cox} and the Behler-Parrinello model, significantly deviate from the expected asymptotic value. 
Consequently, these short-range models are expected to have difficulties describing the dielectric screening that is important in nonuniform systems~\cite{Rodgers:PNAS:2008,Remsing:2016ib,SSM,Cox,niblett2021learning}, for example.

To further examine the dielectric properties of the NN models, we can apply homogeneous fields of varying strength to the system and examine its response. 
To do so, we performed finite-field simulations at constant displacement field, $\vec{D}$. 
These finite-$\vec{D}$ simulations~\cite{Zhang2016} can be naturally combined with our SCFNN model, unlike many other neural network models.
Following previous work~\cite{Cox}, we use $\vec{D}=D\hat{\vec{z}}$, vary the magnitude of the displacement field from $D=0$~V/\AA \ to $D=0.4$~V/\AA, and examine the polarization, $P$, induced in water. 
As shown in Fig.~\ref{fig:dipole}b, the polarization response of water to the external field is accurately predicted by dielectric continuum theory, as expected, further suggesting that the SCFNN model properly describes the dielectric response of water. 
To the best of our knowledge, this is the first NN model that can accurately describe the response of a system to external fields. 
We emphasize that this response is achieved by learning the long-range response via Networks~1L and~2L. 

\section{Discussion}

%
In this work, we have presented a general strategy to construct NN potentials that can properly account for the long-range response of molecular systems that is responsible for dielectric screening and related phenomena. 
We demonstrated that this model produces the correct long-range polarization correlations in liquid water, as well as the correct response of liquid water to external electrostatic fields. 
Both of these quantities are related to the dielectric constant and require a proper description of long-range interactions. 
In contrast, current derivations of NN potentials result in short-range models that cannot capture these effects. 

We anticipate that this approach will be of broad use to the molecular machine learning and simulation community
for modeling electrostatic and dielectric properties of molecular systems. 
In contrast to short-range interactions that must be properly learned to describe the different local environments encountered at extended interfaces and at solute surfaces, the response of the system to long-range, slowly-varying fields is quite general. 
Learning the long-range response (through Networks 1L and 2L) is analogous to learning a linear response, and we expect the resulting model to be relatively transferable. 
As such, our resulting SCFNN model can make \emph{predictions} about conditions on which it was not trained. 
For example, we trained the model for \emph{electric fields} of magnitude 0, 0.1, and 0.2~V/\AA,
and then used this model to successfully predict the response of the system to \emph{displacement fields} with magnitudes between 0~and~0.4~V/\AA. 
This suggests that our approach can be used to train NN models in more complex environments, like water at extended interfaces, and then accurately predict the response of water to long-range fields in those environments.  

Finally, we note that our SCFNN approach is complementary to many established methods for creating NN potentials.
Learning the short-range, GT system interactions can be accomplished with any method that uses local geometric information.
In this case, the precise form of Networks 1S and 2S can be replaced with an alternative NN. 
Then, Networks 1L and 2L can be used as defined here, within the general SCFNN workflow, resulting in a variant of the desired NN potential that can describe the effects of long-range interactions. 
Because of this, we expect our SCFNN approach to be transferable and readily interfaced with current and future machine learning methods for modeling short-range molecular interactions.

\section{Methods}

\subsection{Neural Network Potentials.}
Our training and test set consists of 1571 configurations of 64 water molecules.
Homogeneous electric fields were applied to the system, as described further in the next section. 
We used two thirds of the configurations for training and one third to test the training of the network. 

To train the networks we need to separate the DFT data into the GT system and the long-range effective field.
However, that separation is not straightforward in practice.
To achieve this, we use the differences in the MLWFC locations and forces induced by different fields to fit Networks 1L and 2L.
We now describe this procedure in detail for fitting Network 1L, and Network 2L was fit following a similar approach.

To learn the effects of long-range interactions, we consider perturbations to the positions of the MLWFCs induced by external electric fields of different magnitudes. 
Consider applying two fields of strength $\len{\vec{E}}$ and $\len{\vec{E'}}$.
These fields will alter the MLWFC positions by $\Delta \vec{r}_{w}[\vec{R}, \vec{E}]$ and $\Delta \pvec{r}'_{w}[\vec{R}, \vec{E'}]$, respectively. 
However, both $\Delta \vec{r}_{w}$ and $\Delta \pvec{r}'_{w}$ are not directly obtainable from a single DFT calculation. 
Instead, we can readily compute the difference in perturbations, $\Delta \vec{r}_{w} -\Delta \pvec{r}'_{w} $, directly  from the DFT data, because
 \begin{equation}
 \Delta \vec{r}_{w} -\Delta \pvec{r}'_{w} = \vec{r}_{w} -  \pvec{r}'_{w} \,.
 \end{equation}
 Here, $\vec{r}_{w} $ and $ \pvec{r}'_{w}$ are the locations of the MLWFCs in the full system in the presence of the field $\vec{E}$ and $\vec{E'}$, respectively, and these positions can be readily computed in the simulations. 
 These differences in the MLWFC positions are used to fit Network 1L. 
In addition, we also exploit the fact that $\Delta \vec{r}_w=0$ when $\vec{E}=0$. 
This allows us to fix the zero point of Network 1L. 

 After fitting Networks 1L and 2L, we use them to predict the contribution of the effective field to the MLWFC locations and forces. 
 We then subtract that part from the DFT data. 
 What remains corresponds to the short-range GT system, and this is used to train Networks 1S and 2S.

 We now describe the detailed structure of the four networks used here.
 
\textbf{Network 1S.} In the local frame of water molecule $i$, we construct two types of symmetry functions as inputs to Network 1S.
The first type is the type 2 Behler-Parrinello symmetry function~\cite{Behler2011},
 \begin{equation}
 G_i^2 = \sum_{j \neq i} \exp\para{-\eta(r_{ij}-r_s)^2} f_c(r_{ij}).
 \end{equation}
Here, $\eta$ and $r_s$ are parameters that adjust the width and center of the Gaussian, and $f_c$ is a cutoff function whose value and slope goes to zero at the radial cutoff $r_c$.
We adopted the same cutoff function as previous work~\cite{Morawietz2016}, and the cutoff $r_c$ is set equal to 12~Bohr.

 The second type of symmetry function is similar to the type 4 Behler-Parrinello symmetry function\cite{Behler2011}.
 This symmetry function depends on the angle between $\vec{r}_{ij}$ and the axis of the local frame,
 \begin{equation}
 \vec{G}_i^4 = \sum_{j \neq i} 2^{1-\zeta } (1 + \lambda  \frac{\vec{r}_{ij}}{r_{ij}})^\zeta \exp(-\eta r_{ij}^2) f_c(r_{ij}).
 \end{equation}
 Here,  $\zeta$ and $\lambda$ are parameters that adjust the dependence of the angular term.

We use 36 symmetry functions as input to Network 1S. 
Network 1S itself consists of two hidden layers that contain 24 and 16 nodes. 
The output layer consists of 12 nodes, corresponding to the three-dimensional coordinates of the four MLWFCs of a central water molecule. 
Network 1S is a fully connected feed-forward network, and we use $\tanh(x)$ as its activation function.

 \textbf{Network 1L.} 
 In the local frame of water molecule $i$, we construct one type of symmetry function as input to Network 1L,
  \begin{equation}
 \vec{EG}_i^2 = \sum_{j } \vec{E}_j \exp(-\eta(r_{ij}-r_s)^2) f_c(r_{ij}).
 \end{equation}
 Here, $\vec{E}_j $ is the effective field exerted on atom $j$.
 We use 36 symmetry functions as inputs to Network 1L. 
 Network 1L has no hidden layers. 
 The output layer consists of 12 nodes, corresponding
 to the three-dimensional coordinates of the \emph{perturbations} of a water molecule's four MLWFCs induced by the external field. 
 
\textbf{Network 2S.} Network 2S is exactly the same as the Behler-Parrinello Network employed in previous work~\cite{Morawietz2016}. 
In brief, the network contains 2 hidden layers, each containing 25 nodes. Type 2 and 4 Behler-Parrinello symmetry functions are used as inputs to the network. 
The network for oxygen takes 30 symmetry functions as inputs, while the network for hydrogen takes 27 symmetry functions as inputs.
A hyperbolic tangent is used as the activation function.

\textbf{Network 2L.}
Network 2L uses the same type of symmetry function as Network 1L.
The network for the force on the oxygen and for the force on hydrogen are trained independently. 
To predict the force on the oxygen, we center the local frame on the oxygen atom. 
When the force on a hydrogen atom is the target, we center the local frame on a hydrogen atom. 
We use 36 symmetry functions as inputs to Network 2L. 
Network 2L has no hidden layers. 
The inputs map linearly onto the forces on the atoms.

\subsection{DFT Calculations.}
The DFT calculations followed previous work~\cite{cheng2019ab,marsalek2017quantum} and used published configurations of water as the training set~\cite{cheng2019ab}.
In short, all calculations were performed with CP2K (version 7)~\cite{CP2K,VandeVondele2005}, using
the revPBE0 hybrid functional with 25\% exact exchange~\cite{zhang1998comment,adamo1999toward,goerigk2011thorough}, the D3 dispersion correction of Grimme~\cite{Grimme},
Goedecker-Tetter-Hutter pseudopotentials~\cite{Goedecker1996}, and TZV2P basis sets~\cite{VandeVondele2007}, with a plane wave cutoff of 400~Ry. 
Maximally localized Wannier function centers~\cite{RevModPhys.84.1419} were evaluated with CP2K, using the \texttt{LOCALIZE} option.
The maximally localized Wannier function spreads were minimized according to previous work~\cite{PhysRevB.61.10040}.
A homogeneous, external electric field was applied to the system using the Berry phase approach, with the \texttt{PERIODIC\_EFIELD} option in CP2K~\cite{souza2002first,umari2002ab,zhang2016computing}. 
Electric fields of magnitude 0, 0.1, and 0.2~V/\AA \ were applied along the $z$-direction of the simulation cell.
Sample input files are given at Zenodo (https://doi.org/10.5281/zenodo.5521328).
\subsection{Molecular Dynamics Simulations}
MD simulations are performed in the canonical (NVT) ensemble, with a constant temperature of 300~K maintained using a Berendsen thermostat. 
The system consisted of 1000 water molecules in a cubic box 31.2~\AA \ in length. 
The equations of motion were integrated with a timestep of 0.5~fs. 
Radial distribution functions and longitudinal polarization correlations functions were computed from 100 independent trajectories that were each 50~ps in length. 
Finite-$\vec{D}$ simulations were performed under the same simulation conditions, and each trajectory was 50~ps long at each magnitude of $D$.

\subsection{Data Availability.}
Datasets used to train and test the NNP can be found at Zenodo \\
(https://doi.org/10.5281/zenodo.5521328).

\subsection{Code Availability.}
All DFT calculations were performed with CP2K version 7. 
In-house code was used to construct the NN potentials and perform the MD simulations.
These codes are available at Github (https://github.com/andy90/SCFNN).

\begin{acknowledgements}
We acknowledge the Office of Advanced Research Computing (OARC) at Rutgers,
The State University of New Jersey
for providing access to the Amarel cluster
and associated research computing resources that have contributed to the results reported here.
We thank John Weeks for helpful comments on the manuscript.
\end{acknowledgements}

\bibliography{mainNotes}

\end{document}